\documentclass[12pt]{article}
\usepackage{cite}
\newcommand{\ba}{\begin{array}}
\newcommand{\ea}{\end{array}}
\newcommand{\be}{\begin{equation}}
\newcommand{\ee}{\end{equation}}
\newcommand{\bea}{\begin{eqnarray}}
\newcommand{\eea}{\end{eqnarray}}

\newcommand{\p}{\partial}

\def\IB{\relax\hbox{$\inbar\kern-.3em{\rm B}$}}
\def\IC{\relax\hbox{$\inbar\kern-.3em{\rm C}$}}
\def\ID{\relax\hbox{$\inbar\kern-.3em{\rm D}$}}
\def\IE{\relax\hbox{$\inbar\kern-.3em{\rm E}$}}
\def\IF{\relax\hbox{$\inbar\kern-.3em{\rm F}$}}
\def\IG{\relax\hbox{$\inbar\kern-.3em{\rm G}$}}
\def\IGa{\relax\hbox{${\rm I}\kern-.18em\Gamma$}}
\def\IH{\relax{\rm I\kern-.18em H}}
\def\IK{\relax{\rm I\kern-.18em K}}
\def\IL{\relax{\rm I\kern-.18em L}}
\def\IP{\relax{\rm I\kern-.18em P}}
\def\IR{\relax{\rm I\kern-.18em R}}
\def\IZ{\relax{\rm Z\kern-.5em Z}}

\def\half{\frac{1}{2}}
\def\p{\partial}
\def\f{\frac}

\begin{document}

\begin{titlepage}


\begin{flushright}
IHES/P/01/48  \\
LPT-Orsay-01/108 \\
OUTP-01-38P \\
hep-th/0112008
\end{flushright}

\vskip 2 cm

\begin{center}
{\LARGE Stress Energy tensor in LCFT and the Logarithmic 
Sugawara construction \footnote{This is an expanded version of a talk presented by A. Nichols at the conference on Logarithmic Conformal Field Theory and its Applications in Tehran Iran, 2001}}
\vskip 1 cm

{
\large Ian I.Kogan\footnote{i.kogan@physics.ox.ac.uk, iankogan@ihes.fr, also at ITEP, Moscow, Russia}
$^{\,a,b,c}$
 ~and Alexander Nichols\footnote{a.nichols1@physics.ox.ac.uk} $^{\,a}$
}

\begin{center}
$^a$
{\em  Theoretical Physics, Department of Physics, Oxford University\\
1 Keble Road, Oxford, OX1 3NP, UK } \\
$^b$ 
{\em IHES, 35 route de Chartres,
91440, Bures-sur-Yvette,  France }\\
$^c$
{\em Laboratoire de Physique Th\'eorique,
Universit\'e de Paris XI, \\
91405 Orsay C\'edex, France
}\\
\end{center}

\vskip 1 cm

\vskip .5 cm 

\begin{abstract}
We discuss the partners of the stress energy tensor and their structure in Logarithmic conformal field theories. In particular we draw attention to the fundamental differences between theories with zero and non-zero central charge. However they are both characterised by at least two independent parameters. We show how, by using a generalised Sugawara construction, one can calculate the logarithmic partner of $T$. We show that such a construction works in the $c=-2$ theory using the conformal dimension one primary currents which generate a logarithmic extension of the Kac-Moody algebra.

\end{abstract}

\end{center}

\end{titlepage}

\section{Introduction}

The study of conformal invariance in two dimensions has been an extremely interesting and fruitful area of research for the last twenty years \cite{Belavin:1984vu}. 

During the last ten years an interesting class of conformal field theories (CFTs) has emerged called logarithmic conformal field theories (LCFTs). In \cite{Gurarie:1993xq} the concept of LCFT was introduced and the presence of logarithmic structure in the operator product expansion was explained by the indecomposable representations that can occur in the fusion of primary operators. These occur when there are fields with degenerate scaling dimensions having a Jordan block structure. It was shown that in  any LCFT 
  one of these degenerate fields   becomes a  zero  norm state   coupled to a logarithmic partner
 \cite{Caux:1996nm}. This together with another property - extra (hidden) symmetries   
\cite{Caux:1996nm,Kogan:1996df},  coming from extra conserved currents in our theory, will be important for our analysis of the 
stress-energy tensor structure in LCFT.

LCFTs have emerged in many different areas such as: WZNW models and gravitational dressing \cite{Bilal:1994nx,Caux:1997kq,Giribet:2001qq,Gaberdiel:2001ny,Nichols:2001du}, polymers \cite{Saleur:1992hk,Cardy,Gurarie:1999yx}, disordered systems and the Quantum Hall effect \cite{Caux:1996nm,Kogan:1996wk,Maassarani:1996jn,Gurarie:1997dw,CTT,Caux:1998eu,Bhaseen:1999nm,Kogan:1999hz,Gurarie:1999bp,RezaRahimiTabar:2000qr,Bernard:2000vc,Bhaseen:2000bm,Bhaseen:2000mi,Ludwig:2000em}, string theory 
\cite{Kogan:1996df}, \cite{ Kogan:1996zv}-\cite{Moghimi-Araghi:2001fg},
 2d turbulence \cite{RahimiTabar:1996dh,RahimiTabar:1997nc,Flohr:1996ik,RahimiTabar:1997ki,RahimiTabar:1996si}, multi-colour QCD at low-x 
\cite{Korchemsky:2001nx} and the Seiberg-Witten solution of ${\mathcal N}=2$ SUSY Yang-Mills
\cite{Cappelli:1997qf,Flohr:1998ew}. Deformed LCFTs, Renormalisation group flows and the c-theorem were discussed in  \cite{Caux:1996nm,Rahimi-Tabar:1998ph,Mavromatos:1998sa}. The holographic relation between logarithmic operators and vacuum instability was considered in \cite{Kogan:1998xm,Lewis:1999fg}

There has also been much work on analysing the general structure and consistency of such models in particular the $c_{p,q}$ models and the special case of $c=-2$ which is by far the best understood 
\cite{Kausch:1995py,Gaberdiel:1998ps,Kausch:2000fu,Flohr:2000mc}. It is unclear as yet how much of the structure, for instance the role of extended algebras, is generic to all LCFTs. For more about the general structure of LCFT see \cite{RahimiTabar:1997ub,Rohsiepe:1996qj,Kogan:1997fd,Flohr:2001tj} and references therein. Introductory lecture notes on LCFT and more references can be found in \cite{Tabar:2001et,Flohr:2001zs,Gaberdiel:2001tr}.

One of the interesting features of LCFT is the existence of  the above-mentioned  hidden symmetry  which means there are extra fields with integer conformal dimensions. These may play a role as extensions of the chiral algebra \cite{Hadjiivanov:2001kr}. One can even get extra states with zero dimension which means that we have a theory with a non-trivial vacuum. These operators play a prominent role in the Quantum-Hall effect \cite{Bhaseen:1999nm,Kogan:1999hz}. In this case the descendents of this extra zero dimension operator may form logarithmic pairs with currents or higher dimension fields. It is therefore interesting to see what will happen in the case of the stress tensor itself - can it have logarithmic partners or not? - and will these partners be primary fields or descendents.

In this paper we shall address this issue and try to  suggest some kind of classification for LCFT based on the structure of the vacuum and the character of the degeneracy of the stress-energy tensor. 

In particular the structure of the partners to $T$ in LCFTs with non-zero central charge and LCFTs with zero central charge behave very differently. The second class, $c=0$ theories,  are a very special 
sub-class of LCFTs. They are of utmost importance for both disordered systems  and critical strings.
We shall see that at $c=0$ in order to get a non-trivial theory there must exist a state which is orthogonal to $T$ and is \emph{not} a descendant of any other field. There could of course also be other states which are descendants but these are not required by general arguments. The appearance of such a state is characterised by at least two coefficients. We shall discuss the arguments presented in \cite{Cardy,Gurarie:1999yx,CardyTalk} concerning the existence of a logarithmic partner $t$ for the stress-energy tensor. In particular the emergence of logarithmic behaviour is not universal if we can decompose the theory into a sum of non-interacting sectors. It is the mixing between these sectors which makes the theory logarithmic. This issue was previously discussed in string theory with a ghost-matter mixing term \cite{SUSY30,Kogan:2000nw}.

For $c \ne 0$ we can have a logarithmic partner of $T$ which is a descendent field. There is at least one parameter involved in determining the exact form of this descendent field. However once it is determined the central charge then completely fixes the OPE between $T$ and $t$.

We also discuss a generalisation of the Sugawara construction in logarithmic theories. It will be shown how the presence of extra currents will allow  us to calculate the logarithmic partner $t$. Moreover as we shall demonstrate explicitly in the $c=-2$ model this can be done in a theory which does not have a Kac-Moody symmetry.

\section{Towards the classification of LCFT}
LCFTs can be naturally divided into classes based on the dimension of the Jordan blocks involved. Here we shall concentrate on the case of rank 2 (one logarithmic partner) however it is obvious that our results will generalise to higher rank Jordan cells. It is perhaps still an interesting problem to understand if there can be a more complicated structure at higher rank.

These theories can be further grouped, as we shall explain in this paper, into four distinct categories in which the stress tensor and its partners have different structures:
\begin{itemize}
\item{$c=0$ Theories}
	\begin{itemize}
	\item 0A:~Non-degenerate vacua: ($SU(2)_0$, Disordered Models ??) \\
	\item 0B:~Degenerate vacua ($OSp(2|2)_k$ for certain $k$)
	\end{itemize}
\item{$c \ne 0$ Theories}
	\begin{itemize}
	\item IA:~Non-degenerate vacua \\
	\item IB:~Degenerate vacua ($c_{p,1}$).
	\end{itemize}
\end{itemize}
Throughout this paper we use the notation that non-degenerate and degenerate refer to the single vacuum and the vacuum with logarithmic pair respectively. There may also be other primaries at $h=0$ with a trivial Jordan cell structure and we do not consider this possibility here. We shall also only concentrate on the chiral OPEs and correlators for simplicity. It is however an important and interesting question as to the full non-chiral structure \cite{WorkinProgress}.

Only in the case of the $c_{p,1}$ models and in particular the $c=-2$ triplet model has the structure of the theory been fully elucidated. For the others some of the structure is known from explicit correlation functions. As far as we are aware there has been no examples of type IA in the literature. 
It is easy to see that a logarithmic partner for $T$ can only exist in cases $0A,0B,IB$ by the following simple arguments. 

If $T$ has a logarithmic partner then $T$ itself must be a zero norm state \cite{Caux:1996nm}:
\bea \label{eqn:Tzeronorm}
\left< T(z) T(w) \right> =0
\eea
Now consider the standard OPE for the stress tensor:
\bea
T(z) T(w) \sim \f{c ~I}{2(z-w)^4}+\f{2 T(w)}{(z-w)^2}+\f{\p T(w)}{z-w} + \cdots
\eea
where $I$ is the identity operator. For consistency with (\ref{eqn:Tzeronorm}) we see that we must have:
\bea
c \left< I \right> =c \left< 0|I|0 \right> = c\left< 0|0 \right>=0
\eea
For $c \ne 0$ this implies that the vacuum $\left.|0\right>$ must have zero norm and thus must be part of a logarithmic pair which excludes case IA. Thus partners to the stress tensor $T$ cannot occur in type IA theories. For this reason in the next section when discussing non-degenerate vacua we shall only discuss the case of $c=0$.

\section{Non-degenerate Vacua and $c \rightarrow 0$ limit}

\subsection{$c=0$ Catastrophe}

Here we review the construction given in \cite{Cardy,Gurarie:1999yx,CardyTalk}.
For a primary field of conformal dimension $h$ we use the normalisation:
\bea
\left< V(z) V(0) \right> = \f{A}{z^{2h}}
\eea
Then if we consider the correlator:
\bea
\left< T(z) V(z_1) V(z_2) \right> = \f{A~h}{(z-z_1)^2(z-z_2)^2(z_1-z_2)^{2h-2}}
\eea
The coefficient of the three point function is uniquely fixed by considering the limit $z \rightarrow z_1$ and using the property of a primary field:
\bea
T(z) V(w) \sim \f{h V(w)}{(z-w)^2}+\f{\p V(w)}{z-w} + \cdots
\eea
We now use:
\bea
\left< T(z) T(0) \right> = \f{c}{2 z^4} \left<0|I|0 \right>= \f{c}{2 z^4} ~~~~~~~ \left<0|0 \right> =1 
\eea
and are explicitly using the fact that the identity field has non-zero norm. If $T$ is the only $h=2$ field present in our model we can deduce:
\bea
V(z) V(0) \sim \f{A(c)}{z^{2h}} \left[ 1 +  \f{2h}{c} z^2 T(0) + \cdots \right]
\eea
Clearly for $c=0$ if $A(0) \ne 0$ then the above OPE becomes ill-defined.
However suppose that as $c$ approaches zero there is another field $\tilde{T}$ with dimension $2+\alpha(c)$ that approaches $2$. Then for $c \ne 0$ we have:
\bea \label{eqn:OPE}
V(z) V(0) \sim \f{A(c)}{z^{2h}} \left[ 1 +  \f{2h}{c} z^2 T(0) + 2\tilde{T}z^{2+\alpha(c)} \cdots \right]
\eea
We shall normalise the operator $\tilde{T}$ in the following way:
\bea
\left< \tilde{T}(z) \tilde{T}(0) \right> = \f{1}{c}\f{B(c)}{z^{4+2\alpha(c)}}
\eea
The equation (\ref{eqn:OPE}) now can be expanded for small $c$. We assume here that $A(0)$ is non-zero and discuss exceptions later.
\bea 
V(z) V(0) \sim \f{A(0)}{z^{2h}} + \f{A(0)}{z^{2h}} z^{2+\alpha(c)} \left[ \f{2h}{c} z^{-\alpha(c)} T(0) + 2\tilde{T}  \right] + \cdots \\
 \sim \f{A(0)}{z^{2h}} + \f{A(0)}{z^{2h}} z^{2+\alpha(c)} \left[ -\alpha(c) \f{2h}{c} \ln z T + \f{2h}{c}T + 2\tilde{T} \right] + \cdots 
\eea
As we wish the limit $c \rightarrow 0$ to be well defined we must have a well defined value for:
\bea \label{eqn:bdef}
b^{-1} \equiv -\lim_{c \rightarrow 0} \f{\alpha(c)}{c}=-\alpha'(0)
\eea
If $b$ also vanished then one would have to introduce a third partner to $T$ but this would no longer be a rank $2$ structure.

The other part of the expression must also be regular in the limit $c \rightarrow 0$ and we define this to be proportional to a new field $t$ via:
\bea
\f{h}{b} t=\f{h}{c}T + \tilde{T} ~~~~
\Rightarrow t= \f{b}{c}T+\f{b}{h} \tilde{T}
\eea
We can now calculate the two point function:
\bea
\left< T(z) t(0) \right> &=& \left< T(z) \left[ \f{b}{c} T + \f{b}{h}  \tilde{T} \right](0) \right>  = \f{b}{c} \left< T(z) T(0) \right> \nonumber \\
&=& \f{b}{2 z^4}
\eea
where we used the fact that $\left<T(z) \tilde{T}(0) \right>$ vanishes as they have different dimensions. Also:
\bea
\left< t(z) t(0) \right> &=& \left< \left[ \f{b}{c} T + \f{b}{h}  \tilde{T} \right](z) \left[ \f{b}{c} T + \f{b}{h}  \tilde{T} \right](0) \right> \nonumber \\
&=&\f{b^2}{c^2} \left< T(z) T(0) \right> + \f{b^2}{h^2} \left< \tilde{T}(z) \tilde{T}(0) \right> \nonumber \\
&=&\f{b^2}{2c}\f{1}{z^4}+\f{b^2 B(c)}{h^2 c} \f{1}{z^{4+2\alpha(c)}} \nonumber \\
&=&\f{b^2}{2c}\f{1}{z^4}+\f{b^2 B(c)}{h^2 c} \f{1}{z^4} \left( 1- 2\alpha(c) \ln z + \cdots \right) \\
&=&\f{1}{z^4} \left\{ \left( \f{b^2}{2c}+\f{b^2 B(c)}{h^2 c} \right) - \f{2b^2 B(c)\alpha(c)}{h^2c}\ln z + \cdots \right\} \nonumber
\eea
As this is to be well defined we see from the first part that we must have:
\bea
B(c)=-\f{1}{2} h^2+B_1c +\cdots
\eea
Now using (\ref{eqn:bdef}) the $O(1)$ term becomes $-2b \ln z$. Thus we get the standard OPEs for a logarithmic pair:
\bea
\left< T(z) T(0) \right>&=&0 \nonumber \\
\left< T(z) t(0) \right>&=& \f{b}{2 z^4} \\
\left< t(z) t(0) \right>&=& \f{B_1 -b\ln z}{z^4} \nonumber
\eea
The constant $B_1$ can  be removed by a finite redefinition of $t$ via $t \rightarrow t+\gamma T$. From now on we shall assume that this has been done.
The OPE (\ref{eqn:OPE}) now becomes:
\bea 
V(z) V(0) \sim \f{A(0)}{z^{2h}} \left[ 1 +  \f{2h}{b} z^2 \left( T  \ln z+ t \right) \cdots \right]
\label{logOPE}
\eea
which is now involves quantities that are perfectly well defined in the limit as $c \rightarrow 0$.
We can now continue and insist that $t$ is also well defined in the three point functions (See Appendix). Assuming now that the algebra closes these are then sufficient to determine the OPEs. These are:
\bea
T(z) t(0) &\sim& \f{b}{2 z^4}+\f{2t(0)+T(0)}{z^2}+\f{\p t(0)}{z} +\cdots  \label{OPETt} \\
t(z) t(0) &\sim& \f{-b \ln z}{z^4}+\f{(1-4\ln z)t(0)+(\f{2a}{b}-2\ln^2 z-\ln z)T(0)}{z^2} \nonumber \\
&&~~~+\f{\f{1}{2}(1-4 \ln z)\p t(0)+ \f{1}{2}(\f{2a}{b}- \ln z - 2\ln^2 z)\p T(0)}{z} \cdots  
\label{OPEtt}
\eea
The appearance of a state $\left.|t\right> = t\left.|0\right> $ in this way 
is equivalent to postulating a logarithmic partner for the null vector $T$ \cite{Rohsiepe:1996qj}. This prevents $T$ from decoupling despite the fact that it is a zero-norm state.
Note that once one fixes:
\bea
L_{0}\left.|t\right>=2\left.|t\right>+\left.|T\right>
\eea
then the parameter $b$ cannot be removed by rescaling and thus different values of $b$ correspond to inequivalent representations.

Let us note that in our notation parameter $b$ is different by a factor of 2 from a definition given in 
\cite{Gurarie:1999yx}. We also want to stress that  the $t(z) t(0) $ OPE (\ref{OPEtt})
 is determined by {\bf two} parameters  $a$ and $b$, not by only $b$ as in \cite{Gurarie:1999yx}. 
The constant term  $2a/b$ cannot be removed by scale transformation - one can absorb it in $\ln z$ term, but 
 not into $\ln^2 z$. 
 One of the important open problems  is to  find  if  the classification of all $c=0$ theories of this type can be 
reduced to the classification of all possible pairs $(a,b)$. Let us note that the parameter $a$ cannot be  
 determined from singular terms in  (\ref{logOPE}), but only from (\ref{OPEtt}), i.e. from the full 3-point
 functions of $(T,t)$ pair (see Appendix).

\subsection{$c=0$ and Separability}

There is a third rather trivial way out of the paradox at $c=0$. It is simply that the full theory is constructed from two parts $T=T_1 \oplus T_2,c=c_1+c_2=0$ both having $c_i \ne 0$. Then in the OPE of two fields from one part we will only see the stress tensor for that part rather than the full one. Operators in the full theory are just the direct product $V=V_1 \otimes V_2$. Then:
\bea
V(z) V(0) &=& V_1(z) V_1(0) ~~ V_2(z) V_2(0) \nonumber \\
&\sim&  \f{1}{z^{2h_1}} \left( 1+ z^2 \f{2h_1}{c_1}T_1(0) + \cdots \right)  ~~ \f{1}{z^{2h_1}} \left( 1+ z^2 \f{2h_2}{c_2}T_2(0) + \cdots \right)  \\
&\sim& \f{1}{z^{2h}} \left[ 1+ z^2 \left( \f{2h_1 T_1}{c_1} + \f{2h_2 T_2}{c_2} \right) +\cdots \right] \nonumber 
\eea
This expression is now perfectly well defined as $c_1,c_2 \ne 0$. Of course this is as expected as the two decoupled theories are perfectly regular.

In critical string theory the ghost and matter sectors are normally assumed to be non-interacting. However this may not be the most general if we wish to allow not just positive but also zero norm states in our final theory \cite{SUSY30}.

\section{Degenerate Vacua}

We review here the arguments of \cite{Moghimi-Araghi:2000qn,Moghimi-Araghi:2000dt}. If the identity is degenerate (i.e. has a logarithmic partner) then we have:
\bea
\left< T(z) T(0) \right>=\f{c}{2 z^4} \left< I \right> =0 ~~~~~~~~~ \left< I \right>=0 
\eea
For non-vanishing correlation functions:
\bea \label{eqn:2pt}
\left< V(z) V(0) \right> = \f{1}{z^{2h}}
\eea
then we must have the OPE:
\bea
V(z) V(0) \sim \f{1}{z^{2h}} \left[  (\omega + I \ln z )  + \f{2h}{b} z^2(t+ \ln z T) + \cdots \right]
\eea
The leading terms in the expansion are fixed the form of the two point function (\ref{eqn:2pt}) and by conformal invariance where the logarithmic pair of operators transforms as:
\bea \label{eqn:Tomega}
L_{0}\left.|\omega \right> = \left.|0\right> ~~~~ L_{0}\left.|0\right>=0
\eea
We write $\left.|\omega\right>$ for the state $\left.\omega |0\right>$. The subleading term in the OPE are just the descendents of these primaries:
%
%
%
%
\bea \label{eqn:tdefn}
\left.|T\right> \equiv L_{-2}\left.|0\right> ~~~~ \left.|t\right> \equiv L_{-2}\left.|\omega \right> + \alpha L_{-1}^2 \left.|\omega \right> + \beta  L_{-2}\left.|0\right>
\eea
where $\alpha,\beta$ are arbitrary coefficients. In particular the field $t$ is \emph{not} a primary and is purely a descendent field.

Given (\ref{eqn:tdefn},\ref{eqn:Tomega}) and the standard Virasoro algebra:
\bea \label{eqn:Virasoro}
\left[ L_n, L_m \right] = (n-m) L_{m+n} + \f{c}{12}n(n^2-1) \delta_{n+m,0}
\eea
we can easily deduce:
\bea
L_{0} \left.|t\right>&=&2\left.|t\right>+\left.|T\right> \nonumber \\
L_{1} \left.|t\right>&=&(3+2\alpha) L_{-1}\left.|\omega\right> \\
L_{2} \left.|t\right>&=&(4+6\alpha+\beta \f{c}{2})\left.|0\right> + \f{c}{2}~ \left.|\omega\right> \nonumber
\eea
These are equivalent to the OPE:
\bea \label{eqn:Ttdegen}
T(z) ~t(0) \sim \f{\f{c}{2}\omega + (4+6\alpha+\beta \f{c}{2})I}{(z-w)^4} + \f{(3+2\alpha) \p \omega}{(z-w)^3} + \f{2t+T}{(z-w)^2} + \f{\p t}{(z-w)} + \cdots
\eea
The coefficient $\beta$ can be removed by a simple redefinition of $t$, adding a multiple of $T$. We include it here only for later comparison. The parameter $\alpha$ cannot be removed in such a way. Thus we see that again we have a two parameter family parameterised by $c$ and $\alpha$.

As $t$ is now a descendent all properties of $t$ are reduced, via the Virasoro algebra, to properties of $\omega$ and $\Omega$. Thus we can calculate the two point function:
\bea
\left< T(z) ~t(0) \right> = \f{c}{2(z-w)^4}
\eea
Thus we see that in $c=0$  such a descendent field \emph{cannot} prevent $T$ from decoupling. It is easy to see that in $c=0$ theories no descendent of any fields $\left.|A\right>,\left.|B\right>,\left.|C\right>,...$ will ever prevent $T$ from decoupling. For general $c$ we have:
\bea
\left<T|t\right>&=&\left<L_{2}|\right. \left\{ L_{-1}\left.|A\right> + L_{-2}\left.|B\right> + L_{-3}\left.|C\right> + \cdots \right\} \nonumber \\
&=&\left<0|\right. \left\{ 3L_1 \left.|A\right> + \left(4L_{0}+\f{c}{2} \right) \left.|B\right>+ 5L_{-1}\left.|C\right> + \cdots \right\} \\
&=& \f{c}{2} \left< 0|B \right> \nonumber
\eea
where we do not use any properties of the states  $\left.|A\right>,\left.|B\right>,\left.|C\right>$ only the Virasoro algebra (\ref{eqn:Virasoro}) and that the vacuum satisfies:
\bea
\left<0|\right. L_{n} =0 ~~ n \le 1 
\eea
We thus conclude that in the $c=0$ theories there must be a \emph{primary field} of dimension $2$ which is responsible for the non-decoupling of $T$. Of course there may also be a part of the $t$ field which is a descendent but this is not essential.

A similar argument a $k=0$ shows that for $SU(N)_0$ there must be a primary field with non-vanishing two point function with $J^a$. This is precisely the field $N^a$ found in \cite{Caux:1997kq}.
\bea
J^a(z) N^b(0) \sim \f{\delta^{ab}}{z^2}+\f{i f^{abc}N^c}{z-w}
\eea

\section{Logarithmic Sugawara construction}

As we commented earlier there is another way in which one may avoid the $c=0$ catastrophe. If $A(c)$ in (\ref{eqn:OPE}) also vanishes in the limit then this may cancel the divergence.

This is exactly what occurs in the Kac-Moody theories at level zero. For clarity we shall only discuss here $SU(2)_0$ although arguments are identical for general Lie groups.
\bea
J^a(z) J^b(0) \sim \f{k g^{ab}}{(z-w)^2} + \f{i f^{ab}_c J^c}{z-w} + regular
\eea
The central charge is:
\bea
c=\f{3k}{k+2}
\eea
Thus for small $c$ we have $k=A(c)=2c/3$. The divergent term $1/c$ in front of $T$ in (\ref{eqn:OPE}) is exactly cancelled leaving a finite result. This is expected because the Sugawara stress tensor is still perfectly well defined in these theories.
However in general there will be other operators whose two point function is non-vanishing e.g. $K^a$ \cite{Caux:1997kq}. Then by the previous arguments we must have:
\bea
K^a(z) K^a(0) \sim \f{1}{z^2} \left[ 1 +  \f{2}{b} z^2 (\ln z T(0) + t) + \cdots \right]
\eea
We see that it is possible by examining the $O(1)$ and $O(\ln z)$ terms to read off the operators $T$ and $t$. It then remains to compute their OPE and find the value of $b$ that is realised in this system.

This is what we call the general Logarithmic Sugawara construction. The normal Sugawara construction:
\bea
J^a(z) J^a(0) \sim singular + T + \cdots
\eea
allow us to construct $T$ from primary currents $J^a$. Given primary pre-logarithmic fields $\tilde{J^a}$ we can now produce both $T$ and its logarithmic partner $t$. We must have a $\ln z$ factor in front of $T$ to ensure both sides transform similarly under scaling:
\bea
\tilde{J}^a(z) \tilde{J}^a(0) \sim singular + ( T \ln z + t) \cdots
\eea

Using the free field representation it was shown in \cite{Kogan:2001nj} that although the stress tensor decomposes into two separate parts:
\bea
T_{SU(2)_0}=T_{c=2}+T_{c=-2}
\eea
the correlation functions have a very non-trivial braiding due to the presence of the screening charges in the theory. It is not clear if the operator $t$ in the full theory could also be written as a sum.

In the next section we analyse the bosonic dimension one primary fields that exist in $c=-2$. We shall show that their OPE is of the above form and that we can calculate $t$ from this explicitly.

\section{Logarithmic Sugawara construction in $c=-2$}

\subsection{Symplectic fermions}

We start from the well known symplectic fermion action :
\bea
S= \f{1}{2 \pi} \int d^2 z ~ \left(\p{\phi^2} \bar{\p} \phi^1 + \bar{\p} \phi^2 \p \phi^1 \right)
\eea
Throughout we shall concentrate on the chiral sector of the above fields although it should be noted that since these have a non-trivial zero mode which couples the holomorphic and antiholomorphic parts creating a full non-chiral algebra.
\bea
\phi^1(z) \phi^2(w) \sim -\ln(z-w) ~~~~~ \phi^2(z) \phi^1(w) \sim \ln(z-w) 
\eea

These fields are the extensions of the normal fermionic fields with the extra zero modes included:
\bea
\eta= \p \phi^2 ~~~~ \xi= \phi^1 
\eea
\bea
\xi(z) \eta(w) \sim \f{1}{z-w} ~~~~ \eta(z) \xi(w) \sim \f{1}{z-w} 
\eea
The extra zero modes give us the logarithmic structure. The stress tensor is given by:
\bea \label{eqn:Tc=-2}
T=\p \xi \eta=\p \phi^1 \p \phi^2 
\eea
T generates the standard Virasoro algebra at $c=-2$:
\bea
T(z) T(0) \sim \f{-2}{2 z^4} + \f{2 T(0)}{z^2} + \f{\p T(0)}{z} + \cdots \nonumber
\eea
We now look for all bosonic \emph{primary} fields with $h=1$ (These were noted by Kausch in \cite{Kausch:1995py} and they are the generators of the $SL(2,R)$ symmetry of the action):
\bea
J^A&=&\phi^1 \p \phi^2 + \phi^2 \p \phi^1 \nonumber\\
J^B&=&\phi^1 \p \phi^1 \\
J^C&=&\phi^2 \p \phi^2 \nonumber
\eea
There are no others due to the fermionic nature of $\phi^1$ and $\phi^2$. To ensure closure one must also add $\omega=-\phi^1 \phi^2$.

We have the standard OPEs for the primary fields $J_a$ ($a=A,B,C$):
\bea
T(z) J^a(0) \sim \f{J_a(0)}{z^2} + \f{\p J_a(0)}{z} + \cdots
\eea
The field $\omega$ is the logarithmic partner of the identity and satisfies:
\bea
T(z) \omega(0) \sim \f{1}{z^2} + \f{\p \omega(0)}{z} 
\eea

\subsection{Logarithmic algebra}

%

Calculating the OPEs and rescaling the fields then we find that these satisfy the following algebra:
\bea
J^3=\f{J^A}{3} ~~~~
J^+=\f{2 J^B}{3} ~~~~
J^-=\f{2 J^C}{3} \nonumber
\eea
\bea
J^3(z) J^3(0) &\sim& \f{2}{9} g^{33} \left( \f{\ln z + \omega + 1}{z^2} + \half \f{\p \omega}{z} \right) \nonumber\\
J^3(z) J^{\pm}(0) &\sim& \pm \f{J^{\pm}}{z} \\
J^+(z) J^-(0) &\sim& \f{2}{9} g^{+-} \left( \f{\ln z + \omega + 1}{z^2} + \half \f{\p \omega}{z} \right) - \f{2 J^3}{z}\nonumber
\eea

Where $g^{33}=1,g^{+-}=-2$ is the metric for the $SL(2,R)$ group.

This is now in a very similar form to a normal affine Lie algebra. Our Logarithmic 'Kac-Moody' algebra is:
\bea \label{eqn:LogKM}
J^a(z) J^b(0) &\sim&  k g^{ab} \left( \f{\ln z + \omega + 1}{z^2} + \half \f{\p \omega}{z} \right) + \f{ f^{ab}_c J^c}{z} \nonumber\\
J^a(z) \omega(0) &\sim&  J^a \ln z \\
\omega(z) \omega(0) &\sim& -\ln^2 z - 2 \omega \ln z\nonumber
\eea

The field $\omega$ almost commutes with the currents $J^a$ up to the $\ln z$ terms. An algebra with indecomposable representations was also commented on in \cite{Read:2001pz}.

We can now consider our Logarithmic Sugawara construction using our above algebra by taking the $O(1)$ and $O(\ln z)$ terms in the OPE expansion of $J^a(z) J^a(0)$. One finds:
\bea
J^a(z) J^a(0) &\sim&  k g^{aa} \left( \f{\ln z + \omega + 1}{z^2} + \half \f{\p \omega}{z} + T\ln z + t + \cdots \right) 
\eea
$T=\p \phi^1 \p \phi^2$ is the normal $c=-2$ stress tensor as before (\ref{eqn:Tc=-2}). The other field $t$ is given by:
\bea
t=-\phi^1 \phi^2 \p \phi^1 \p \phi^2 - \f{3}{4} \p^2 \omega -\f{5}{2}T
\eea
This obeys:
\bea \label{eqn:NewTt}
T(z)t(0) \sim \f{-2-\omega}{z^4}- \f{1}{2} \f{\p \omega}{z^3}+\f{2t+T}{z^2}+\f{\p t}{z}
\eea
To make connection with our earlier section we use the general definition of a descendent field:
\bea
L_{-2} \omega(w) &=& \f{1}{2\pi i} \oint_{z=w} \f{1}{z-w} T(z) \omega(w) dz \\
&=& \f{1}{2\pi i} \oint_{z=w} \f{1}{z-w} \left[ \f{1}{(z-w)^2} + \f{\p \omega(w)}{z-w} + \left(\p^2 \omega + 2T - \phi^1 \phi^2 \p \phi^1 \p \phi^2 \right) \right] \nonumber \\
&=& - \phi^1 \phi^2 \p \phi^1 \p \phi^2  + \p^2 \omega + 2T 
\eea
Then:
\bea
 \left.|t \right>=L_{-2} \left.|\omega \right>-\f{7}{4} L_{-1}^2 \left.|\omega \right> -\f{9}{2} L_{-2} \left.|0 \right>
\eea
Therefore the free parameters in our previous expression (\ref{eqn:Ttdegen}) are fixed in this case to be $\alpha=-\f{7}{4}, \beta=-\f{9}{2}$. Using these values we see that we have complete agreement between (\ref{eqn:Ttdegen}) and (\ref{eqn:NewTt}). As we stated previously $\beta$ can be absorbed into a redefinition of $t$ but $\alpha$ cannot be.

\subsection{Comparison with $W_3$ algebra}

The $c=-2$ theory is normally classified using the $W(2,3^3)$ algebra with generators \cite{Kausch:1995py}:
\bea
W^+&=&\p^2 \phi^2 \p \phi^2 \nonumber\\
W^3&=& \half \left( \p^2 \phi^2 \p \phi^1 + \p^2 \phi^1 \p \phi^2 \right) \\
W^-&=&\p^2 \phi^1 \p \phi^1\nonumber
\eea
To ensure closure one must add the stress tensor $T=\p \phi^1 \p \phi^2$.

We get the algebra:
{\small
\bea
T(z) T(w) &\sim& \f{-2}{2 (z-w)^4} + \f{2 T(w)}{(z-w)^2} + \f{\p T(w)}{z-w} \nonumber \\
T(z) W^a(w) &\sim& \f{3 W^a(w)}{(z-w)^2}+ \f{\p W^a}{z-w} \\
W^a(z) W^b(w) &\sim&  g^{ab} \left( \f{1}{(z-w)^6}- 3 \f{T(w)}{(z-w)^4}-\f{3}{2} \f{\p T(w)}{(z-w)^3} + \f{3}{2} \f{\p^2 T(w)}{(z-w)^2} -4 \f{(T^2)(w)}{(z-w)^2} \right. \nonumber  \\
&&~~~~ \left. + \f{1}{6} \f{\p^3 T(w)}{z-w} -4 \f{\p(T^2)(w)}{z-w} \right) \nonumber \\ 
&& - 5 f^{ab}_c \left( \f{W^c(w)}{(z-w)^3} + \f{1}{2}\f{\p W^a(w)}{(z-w)^2} + \f{1}{25} \f{\p^2 W^c(w)}{z-w} + \f{1}{25} \f{(TW^c)(w)}{z-w} \right) \nonumber
\eea}

This algebra involves quadratic expressions and is not of a simple linear form. In particular it is only associative if certain null states decouple from the theory \cite{Gaberdiel:1996np}. We do not presently know if there are similarly restrictions are on the value of $k$ in (\ref{eqn:LogKM}) and on the representations allowed. As this is not a chiral algebra the associativity cannot be studied from the Jacobi identity and one must instead use the equivalent statement in terms of the crossing symmetry of the four point functions.

Our dimension one fields that we have considered fields are given by $J^a=W^a_{-1}\omega$. The above fields are also related to our earlier 'Logarithmic Kac-Moody algebra' by the substitution $\phi \rightarrow \p \phi$. The currents $J^a$ become the $W_3$ triplet and the field $\omega$ becomes the stress tensor $T$.

So far we have only discussed the bosonic primary fields. However there are also fermionic $h=1$ fields given by $\p \phi_1,\p \phi_2$. Closure of the algebra implies that there should also be $h=0$ fields $\phi_1,\phi_2$. In fact such an extension of the $W(2,3^3)$ algebra is also possible by adding the chiral symplectic fermionic fields at $h=1$ $\p \phi^1, \p \phi^2 $. In this way we obtain the maximal extension of the $W$ algebra. This may shed some light on the fact that the symplectic fermion theory was found to be the only local theory formed from the triplet model \cite{Gaberdiel:1998ps}.

\section{Conclusions}

We have shown that there are several very different structures for the stress tensor and its partners in LCFT. In particular in non-trivial $c=0$ theories we have demonstrated that it is necessary for there to be a primary field of dimension $2$ orthogonal to the stress tensor. The indecomposable representations are characterised by at least two parameters: $a$ and $b$. For $c \ne 0$ in order to have a logarithmic partner $t$ to the stress tensor we cannot have a non-degenerate vacuum state. In this case $t$ can be a simple descendent and in this way all the OPEs of $T$ and $t$ are fully determined. In this process there are again at least two coefficients: the central charge $c$ and another parameter $\alpha$. We have shown that extra conserved currents present in LCFT can be used to generate these extra fields. We have explicitly demonstrated this in the case of the symplectic fermion model at $c=-2$ using a Logarithmic 'Kac-Moody' algebra generated by the conserved currents.

There are clearly many further points that we would like to understand. On a technical level it is currently not possible to write mode expansions for these fields and use the Jacobi identity to check associativity and null vector decoupling conditions. One can also however study associativity from the consistency of the four point functions which should be perfectly feasible. It would be interesting to understand if such non-chiral algebras can be used to classify states of the theory as this could offer a different point of view of theories which are inherently non-chiral. On a more conceptual level we know that $T$ generates conformal transformations but it is unclear if such a simple description can be found for $t$. It would be interesting to investigate coset constructions with logarithmic current algebras. It may also be possible to derive them  from a WZNW type action.

\section{Acknowledgements}
A.N would like to thank the organisers of the LCFT conference in Tehran, Iran for their very great hospitality in particular S. Rouhani, M. Reza Rahimi Tabar and S. Davarpanah. A.N is also very grateful to M. Flohr, M. Gaberdiel, Y. Ishimoto,  S. Kawai and S. Moghimi-Araghi for interesting and stimulating discussions at this conference. We would both like to thank M.J. Bhaseen for general discussions as well as M. Flohr, V. Gurarie and J. Cardy for comments in the final proofreading. A.N is funded by the Martin Senior Scholarship, Worcester College, Oxford. I.I.K is partly supported by PPARC rolling grant PPA/G/O/1998/00567 and EC TMR grant HPRN-CT-1999-00161.

\section{Appendix: Operator product expansions}

Our starting point will be the 3-point functions for $c \ne 0$:
\bea 
\left< T(z_1) T(z_2) T(z_3) \right> &=&\f{c}{z_{12}^2 z_{13}^2 z_{23}^2 } \\
\left< T(z_1) T(z_2) \tilde{T}(z_3) \right> &=&0 \\
\label{eqn:OPEs}
\left< T(z_1) \tilde{T}(z_2) \tilde{T}(z_3) \right> &=&\f{1}{c}\f{C(c)}{z_{12}^2 z_{13}^2 z_{23}^{2+2\alpha(c)} }\\
\left< \tilde{T}(z_1) \tilde{T}(z_2) \tilde{T}(z_3) \right> &=&\f{1}{c^2}\f{D(c)}{z_{12}^{2+\alpha(c)} z_{13}^{2+\alpha(c)} z_{23}^{2+\alpha(c)} }
\eea
The second equation is deduced from assuming a general form and then taking the limit $z_1 \rightarrow z_2$ and using the fact that $T$ and $\tilde{T}$ have different dimensions for $c \ne 0$ and so $\left< T \tilde{T} \right>=0$.

As we wish to have well defined operators $T,t$ they must in particular have regular 3-point functions. This will be enough to determine the leading behaviour of $C(c)$ and $D(c)$.

In the limit $c \rightarrow 0$ we get:
\bea
\left< T(z_1) T(z_2) T(z_3) \right> =0 
\eea
Now consider:
\bea
\left< T(z_1) T(z_2) t(z_3) \right> &=& \left< T(z_1) T(z_2) \left[ \f{b}{c}T+\f{b}{h} \tilde{T} \right] (z_3) \right> \nonumber \\
&=& \f{b}{z_{12}^2 z_{13}^2 z_{23}^2 }
\eea
Also:
\bea
\left< T(z_1) t(z_2) t(z_3) \right> &=& \left< T(z_1)  \left[ \f{b}{c}T+\f{b}{h} \tilde{T} \right] (z_2) \left[ \f{b}{c}T+\f{b}{h} \tilde{T} \right] (z_3) \right> \nonumber \\
&=& \f{b^2}{c^2}\f{c}{z_{12}^2 z_{13}^2 z_{23}^2 } +\f{b^2}{h^2 c}\f{C(c)}{z_{12}^2 z_{13}^2 z_{23}^{2+2\alpha(c)} }  \\
&=& \f{b^2}{z_{12}^2 z_{13}^2 z_{23}^2 } \left[ \f{1}{c} + \f{C(c)}{h^2 c}\left(1-2\alpha(c) \ln z_{23} +\cdots \right) \right] \nonumber
\eea
Thus we must have $C(c)=-h^2 + \cdots $ for regular behaviour as $c \rightarrow 0$.
\bea
C(c)=-h^2+C_1c+C_2c^2+\cdots
\eea

Expanding (\ref{eqn:OPEs}) in the limit $z_1 \rightarrow z_2$ we see:
\bea
(2+\alpha(c)) \left< \tilde{T}(z) \tilde{T}(0) \right> = \f{1}{c}\f{C(c)}{z^{4+2\alpha(c)}}\\
\Rightarrow (2+\alpha(c))B(c)=C(c)
\eea
Thus:
\bea
(2+\alpha(c))\left(-\f{h^2}{2}+B_1c + O(c^2) \right)=-h^2+C_1c+O(c^2) \\
\Rightarrow -h^2+\left(2B_1+\f{h^2}{2b} \right)c +O(c^2)=-h^2+C_1c+O(c^2)
\eea
As expected the $O(1)$ terms agree. As explained in the text in deriving the two point functions we removed the constant $B_1$ by a redefinition of $t$. For consistency we must therefore have:
\bea
C_1=\f{h^2}{2b}
\eea

Thus
\bea
\left< T(z_1) t(z_2) t(z_3) \right> = \f{-2b\ln z_{23}+\f{C_1b^2}{h^2}}{z_{12}^2 z_{13}^2 z_{23}^2 }=\f{-2b\ln z_{23}+\f{b}{2}}{z_{12}^2 z_{13}^2 z_{23}^2 }
\eea
Finally:
\bea
\left< t(z_1) t(z_2) t(z_3) \right>  &=& \left<  \left[ \f{b}{c}T+\f{b}{h} \tilde{T} \right] (z_1)   \left[ \f{b}{c}T+\f{b}{h} \tilde{T} \right] (z_2) \left[ \f{b}{c}T+\f{b}{h} \tilde{T} \right] (z_3) \right>  \nonumber  \\
&=& \f{b^3}{c^3} \left< T(z_1) T(z_2) T(z_3) \right> + \f{b^3}{h c} \left( \left< \tilde{T}(z_1) T(z_2) T(z_3) \right> \right.  \nonumber  \\
&&\left. + \left< T(z_1) \tilde{T}(z_2) T(z_3) \right> + \left< T(z_1) T(z_2) \tilde{T}(z_3) \right> \right) +\f{b^3}{c^3} \left< \tilde{T}(z_1) \tilde{T}(z_2) \tilde{T}(z_3) \right>  \nonumber  \\
&=&\f{b^3}{c^2} \f{1}{z_{12}^2 z_{13}^2 z_{23}^2 } + \f{b^3}{h c^2} \f{C(c)}{z_{12}^2 z_{13}^2 z_{23}^2 } \left[ z_{12}^{-\alpha(c)} + z_{13}^{-\alpha(c)} z_{23}^{-\alpha(c)} \right] \\
&&+ \f{b^3}{h^3 c^2} \f{D(c)}{z_{12}^2 z_{13}^2 z_{23}^2 } z_{12}^{-\alpha(c)} z_{13}^{-\alpha(c)} z_{23}^{-\alpha(c)}  \nonumber 
\eea
Now expanding this and using the fact that $\alpha(c)$ is $O(c)$ and $\alpha'(0)=-1/b$ we find:
\bea
\left< t(z_1) t(z_2) t(z_3) \right>  &=& \f{b^3}{h^3 c^2} (-2h^3+D_0) \\
&&+ \f{b^2}{h^3 c} \left( \left(D_0-2h^3)(\ln z_{12}+ \ln z_{13}+ \ln z_{23} \right) + 3C_1 hb+D_1b \right) +O(1)  \nonumber 
\eea
Thus if this is to be regular in the limit we must have:
\bea
D_0=2h^3 ~~~~~ D_1=-3C_1 h=-\f{3h^3}{2b}
\eea
Then from the $O(1)$ terms we get:
\bea
\left< t(z_1) t(z_2) t(z_3) \right>  &=&\f{1}{z_{12}^2 z_{13}^2 z_{23}^2 } \Biggl\{ -b\left( \ln^2 z_{12} +  \ln^2 z_{13} +  \ln^2 z_{23} \right)   \nonumber  \\
&&+ 2b \left(\ln z_{12} \ln z_{13} + \ln z_{12} \ln z_{23} + \ln z_{13} \ln z_{23} \right)  \\
&& - \f{b}{2} \left( \ln z_{12} +  \ln z_{13} +  \ln z_{23}  \right) + a \Biggr\}  \nonumber
\eea
where we have defined the constant $a$ by:
\bea
a \equiv \f{b^3}{h^3}\left(D_2 +3C_2h \right)
\eea

Expanding these in the limit we get the OPEs stated in the text.




\end{document}